\documentclass[useAMS,usenatbib]{mn2e}
\usepackage{hyperref}
\usepackage{graphicx}
\usepackage{amsmath}
\usepackage{amssymb}

\title[Escape from the vicinity of fractal basin boundaries of a star cluster]{Escape from the vicinity of fractal basin boundaries of a star cluster}
\author[A. Ernst, A. Just, R. Spurzem \& O. Porth]{A. Ernst$^{1,2}$\thanks{email: aernst@ari.uni-heidelberg.de}, 
A. Just$^{1}$\thanks{email: just@ari.uni-heidelberg.de},
R. Spurzem$^{1}$\thanks{email: spurzem@ari.uni-heidelberg.de} and
O. Porth$^1$\thanks{email: oporth@ari.uni-heidelberg.de}\\
\\
$^{1}$Astronomisches Rechen-Institut/Zentrum f\"ur Astronomie der Universit\"at Heidelberg, 
M\"onchhofstrasse 12-14,
69120 Heidelberg, Germany\\
$^{2}$Max-Planck-Institut f\"ur Astronomie, K\"onigstuhl 17, 69117 Heidelberg, Germany}

\begin{document}

\date{Accepted ... Received ...}

\pagerange{\pageref{firstpage}--\pageref{lastpage}} \pubyear{2002}

\maketitle

\label{firstpage}

\begin{abstract}
The dissolution process of star clusters is rather intricate for theory. 
We investigate it in the context of chaotic dynamics.  We use the simple Plummer model for the gravitational field of a star cluster and treat the tidal field of the Galaxy within the tidal 
approximation. That is, a linear approximation of tidal forces from the Galaxy based on epicyclic 
theory in a rotating reference frame. 
The Poincar\'e surfaces of section reveal the effect of a Coriolis asymmetry. 
The system is non-hyperbolic which has important
consequences for the dynamics. We calculated the basins of escape
with respect to the Lagrangian points $L_1$ and $L_2$. The longest escape times 
have been measured for initial conditions in the vicinity of the fractal basin boundaries.
Furthermore, we computed the chaotic saddle for the system and its stable
and unstable manifolds. The chaotic saddle is a fractal structure in phase space which 
has the form of a Cantor set and introduces chaos into the system. 
\end{abstract}


\raggedbottom

\section{Introduction}

The dissolution process of star clusters is an old problem in stellar dynamics.
Once a star cluster has formed somewhere in a galaxy, it tends to lose mass 
due to dynamical interactions until it has completely dissolved. 
It turned out that the physics behind the dissolution of star clusters is 
intricate and fascinating for theory.
If a star cluster of finite mass were isolated, in virial equilibrium (i.e. $\langle v_e^2 \rangle = 12\sigma_{1D}^2$)
and the velocity distribution given by a Maxwellian
$f_M(X) = \left(4/\sqrt{\pi}\right) X^2 \exp(-X^2)$, where  $X=v/(\sqrt{2}\sigma_{\rm 1D})$ and 
$v, v_e$ and $\sigma_{1D}$ are the velocity, the escape velocity and the velocity dispersion, 
respectively, the fraction of stars which are faster than the rms escape speed
were given by $\chi_e = \int_{\sqrt{6}}^\infty f_M(X) dX = 2\sqrt{6/\pi} \exp(-6) + {\rm erfc}\sqrt{6} \simeq 0.00738316$. This simple analytical result was published by Ambartsumian (1938) and two 
years later, independantly by Spitzer (1940) who named this effect ``evaporation''.
The relevant process which brings stars above the escape speed and lets them evaporate, is 
two-body relaxation. The time scale of relaxation, which determines the rate of dynamical evolution of a star cluster, yields thus an 
upper limit to the lifetime of any star cluster. However, since $\chi_e$ is so small (and
relaxation time relatively long), the 
evaporation time is much longer than a Hubble time for typical globular clusters. 
Following a suggestion of Chandrasekhar (1942),  King (1959) studied the effect of 
 ``potential escapers''. These are stars which have been scattered above 
the escape energy but which have not yet left the cluster. These
may be scattered back to negative energies within a crossing time and remain
bound.  H\'enon (1960, 1969) stressed the importance of few close encounters between stars 
for the rate of mass loss of star clusters.  The Fokker-Planck approximation, which is widely used
to study the dynamical evolution of star clusters, neglects strong encounters by construction.
Nevertheless, close encounters could still be interpreted statistically as a certain discontinuous 
Markov process (Tscharnuter 1971). However, direct $N$-body models seem to be ideally 
suited to study this phenomenon in more detail. Spitzer \& Shapiro (1972) estimate that `` 
close encounters may produce effects perhaps as great as 10 percent of the ``dominant'' 
distant encounters'' and ignore them. 
%safely.

Nature provides an environment for star clusters in which the escape rate is typically strongly enhanced 
as compared with the slow evaporation rate of isolated star clusters: The tidal field of a galaxy 
induces saddle-like troughs in the walls of the potential well of a star cluster (cf. Figure 1). 
It therefore lowers the energy threshold in star clusters above which stars can escape from zero 
to a negative value (Wielen 1972, 1974). 
Moreover, if we consider a star cluster in the tidal field of a galaxy as a dynamical system, 
the tidal field can change the system's dynamics in a dramatical way 
as compared with an isolated system.

In general, the escape process from star clusters in a tidal field 
proceeds in two stages: (1) Scattering of stars into the ``escaping phase 
space'' by two-body encounters and (2) leakage through 
openings in the equipotential sufaces around saddle points of the potential.
The ``escaping phase space'' is defined as the subset of phase space, from which escape is 
possible.  It is well understood that the time scale for a star to complete stage (1) scales 
with relaxation time. On the other hand, the time scale for a star to complete stage (2) depends mainly 
on its energy (but also on its location in phase space as we will see). When we neglect the effect of two-body relaxation for the consideration 
of stage (2), the motion  of a single star in the star cluster is determined between times $t_1$ and 
$t_2$ only by the smooth gravitational potential in which 
the star moves. The potential itself is generated by the other stars in the star cluster disregarding their 
``grainyness'' and by the superposed galactic gravitational field, which is due to the matter 
distribution of the galaxy. Within this framework, we will study stage (2) of the escape process 
in this paper. In this connection, the work of Fukushige \& Heggie (2000) is of major
interest. Their main result is an expression for the time scale of escape
for a star in stage (2) which has just completed stage (1). The dependance of the escape process
on two time scales which scale differently with the particle number $N$ imposes
a severe scaling problem for $N$-body simulations. The scaling problem is of relevance
since the it is on today's general-purpose hardware architectures not yet simply feasible
to simulate the evolution of globular clusters with realistic particle numbers of a few hundred
thousands or even millions of stars by means of direct $N$-body simulations. 
The result of Fukushige \& Heggie has been applied in Baumgardt (2001) to solve the 
important scaling problem for the dissolution time of star clusters in the special case of
circular cluster orbits. The obtained scaling law $t_{dis}\propto t_{rh}^{3/4}$, where
$t_{dis}$ and $t_{rh}$ are the dissolution and half-mass relaxation times, respectively has been verified, e.g. in Spurzem et al. (2005).

The problem of escape has also a long history in the context of the theories of dynamical systems 
and chaos. It is well-known for a long time, that certain Hamiltonian systems allow for escape 
of particles towards infinity. Such
``open'' Hamiltonian systems have been studied by Rod (1973), Churchill et al. (1975),
Contopoulos (1990), Contopoulos \& Kaufmann (1992), Siopis et al. (1997), 
Navarro \& Henrard (2001) and Schneider, T\'el \& Neufeld (2002). 
The related chaotic scattering process, in which a particle
approaches a dynamical system from infinity, interacts with the system and leaves it,
escaping to infinity, was investigated by many authors, as Eckhardt \& Jung (1986), Jung (1987),
Jung \& Scholz (1987), Eckhardt (1987), Jung \& Pott (1989), Bleher, Ott \& Grebogi (1989) and
Jung \& Ziemniak (1992). Chaotic scattering
in the restricted three-body problem has been studied by Benet et al. (1997, 1999).
Typically, the infinity acts as an attractor for an escaping particle, which may escape through
different exits in the equipotential surfaces. Thus it is possible
to obtain basins of escape (or ``exit'' basins), similar to basins of attraction in dissipative
systems or the well-known Newton-Raphson fractals. Special types of basins of attraction
(i.e. ``riddled'' or ``intermingled'' basins)  have been explored by Ott et al. (1993) and 
Sommerer \& Ott (1993, 1996). Basins of escape have been studied by Bleher et al. (1988), and 
they are discussed in Contopoulos (2002).
Reasearch on escape from the paradigmatic H\'enon-Heiles system has been done by de 
Moura \& Letelier (1999), Aguirre, Vallejo \& Sanju\'an (2001), Aguirre \& Sanju\'an (2003), 
Aguirre, Vallejo \& Sanju\'an (2003), Aguirre (2004) and Seoane Seoane Sep\'ulveda (2007). 
These papers served as the basis of this work. Relatively early, it was recognized,
that the key to the understanding of the the chaotic scattering process is a fractal structure in
phase space which has the form of a Cantor set (Cantor 1884) and is called the chaotic saddle. Its skeleton consists of unstable periodic orbits (of any period) which are dense on the chaotic
saddle (e.g. Lai 1997) and introduce chaos into the system (e.g. Contopoulos 2002).
The properties of chaotic saddles have been investigated by different authors, as
Hunt (1996), Lai et al. (1993), Lai (1997) or Motter \& Lai (2001). Both hyperbolic and non-hyperbolic chaotic saddles occur
in dynamical systems. In the first case, there are no Kolmogorov-Arnold-Moser (KAM) tori, which 
means that all periodic orbits are unstable. In the second case, there are both KAM tori and chaotic sets in the phase space (J. C. Vallejo, priv. comm. and e.g. Lai et al. 1993). We note that all of the above 
references on the chaotic dynamics are exemplary rather than exhaustive since there exists 
a vast amount of literature on these topics.

The aim of this paper is to allude to the importance of this last-mentioned branch of 
research for the field of stellar dynamics of star clusters. We will study the escape process 
from star clusters within the framework of chaotic dynamics. In section 2, we introduce the tidal
approximation, i.e. approximate equations of motion for stellar orbits in a
star cluster which is embedded in the tidal field of a galaxy. In section 3, we describe
our (very simple) model of the gravitational potential. In section 4, we discuss
Poincar\'e surfaces of section, which show the effect of a Coriolis asymmetry. Furthermore,
we discuss the basins of escape in Section 5 and the chaotic saddle and its stable and unstable
invariant manifolds in Section 6. Section 7 contains the discussion and conclusions.

\section{The tidal approximation}

The ``tidal approximation''  which is widely used in stellar and galactic dynamics for studies
of stellar systems in a tidal field is nothing else than a simple approximation which, historically, 
has been applied already in the $19$th century in ``Hill's problem'' (e.g. Stumpff 1965, 
Szebehely 1967, Siegel \& Moser 1971) in the context of the (rather intricate) 
lunar theory. The difference to the tidal approximation 
lies merely in the form of the gravitational potentials which are used.
The assumption that a star cluster moves around the Galactic centre 
on a circular orbit allows to use the epicyclic approximation to
calculate steady linear tidal forces acting on the stars in the star cluster. 
As in the circular restricted three-body problem the appropriate coordinate system 
is a rotating reference frame in which both the star cluster centre and the Galactic centre 
(i.e., the primaries) are at rest. Its origin is the star cluster centre, sitting in the minimum of 
the effective Galactic potential. 
The x-axis points away from the Galactic centre; the y-axis points in the 
direction of the rotation of the star cluster around the Galactic centre; the z-axis lies 
perpendicular to the orbital plane and points towards the Galactic North pole. 
We define scaled corotating coordinates $(x,y,z)$  as

\begin{equation}
x = \left(R - R_g\right)/r_t, \ \ \ \ \ y \simeq R_g \left(\phi - \omega t\right)/r_t, \ \ \ \ \ z = z'/r_t
\end{equation}

\noindent
where $(R, \phi, z')$ are galactocentric cylindrical coordinates, $R_g$ and $\omega$ are the 
radius and the frequency of the circular orbit, respectively, $r_t$ is a length scale (i.e. the tidal
radius defined in Equation (\ref{eq:rtidal}) below) and $t$ is time (in the
context of ``Hill's problem'' cf. Glaschke 2006). 
Since the coordinate system is rotating, centrifugal and Coriolis forces appear
according to classical mechanics. In addition, tidal forces enter
the equations of motion for stellar orbits near the origin of coordinates. 
To first order, we have in the rotating frame

\begin{eqnarray}
\ddot{x} &=& -\frac{\partial \Phi_{cl}}{\partial x} - \left(\frac{\partial^2 \Phi_g}{\partial R^2}\right)_{(R_g, 0)} x + \omega^2 x + 2\omega \dot{y} \label{eq:eqm1b} \\
\ddot{y} &=& -\frac{\partial \Phi_{cl}}{\partial y} - 2\omega\dot{x} \label{eq:eqm2b}\\
\ddot{z} &=& -\frac{\partial \Phi_{cl}}{\partial z} -\left(\frac{\partial^2 \Phi_g}{\partial z^2}\right)_{(R_g, 0)} z \label{eq:eqm3b}
\end{eqnarray}

\noindent
where $\Phi_{cl}(x,y,z)$ and $\Phi_g(R,z')$ are the star cluster potential and the axisymmetric
galactic potential, respectively. The second-last term on the right hand side in (\ref{eq:eqm1b}) is the centrifugal force and the last terms in (\ref{eq:eqm1b}) and (\ref{eq:eqm2b}) are Coriolis forces.
According to Binney \& Tremaine (1987), the epicyclic frequency $\kappa$ and the vertical 
frequency $\nu$ are given by

\begin{equation}
\kappa^2 = \left(\frac{\partial^2 \Phi_g}{\partial R^2}\right)_{(R_g,0)} + 3\omega^2, \ \ \ \ \ \ 
\nu^2 = \left(\frac{\partial^2 \Phi_g}{\partial z^2}\right)_{(R_g, 0)} 
\end{equation}

\noindent
Thus the equations of motion can be written as

\vspace{-0.3cm}

\begin{eqnarray}
\ddot{x}&=&f_x - (\kappa^2 - 4\omega^2) x + 2\omega \dot{y} \label{eq:eqm1} \\
\ddot{y}&=&f_y - 2\omega\dot{x} \label{eq:eqm2} \\
\ddot{z}&=&f_z - \nu^2 z, \label{eq:eqm3} 
\end{eqnarray}

\noindent
where $(f_x, f_y, f_z) = -\nabla\Phi_{\mathrm{cl}}$ is the (specific) force vector from the other cluster member stars which typically depends non-linearly on the coordinates.  
It is of interest for the following discussion that the equations of motion
(\ref{eq:eqm1}) - (\ref{eq:eqm3}) are invariant under time reversal.\footnote{The invariance
under time reversal is related to the existence of a discrete group with only two elements, 
which acts on the space of solutions of the equations of motion (\ref{eq:eqm1}) - (\ref{eq:eqm3}).
In our case, the effective potential (\ref{eq:effpot}) and the Coriolis forces are time-symmetric,
which implies the same symmetry of the equations of motion.}
Note that under a time reversal the frequencies also change their sign.
Also, the equations of motion (\ref{eq:eqm1}) - (\ref{eq:eqm3}) admit an isolating integral 
of motion, the Jacobian

\begin{equation}
C=\frac{1}{2} \left(\dot{x}^2 + \dot{y}^2 + \dot{z}^2\right)  + \Phi_{\rm eff}(x,y,z), \label{eq:jacobian} \\
\end{equation}

\noindent
where

\begin{figure*}
\includegraphics[width=0.9\textwidth]{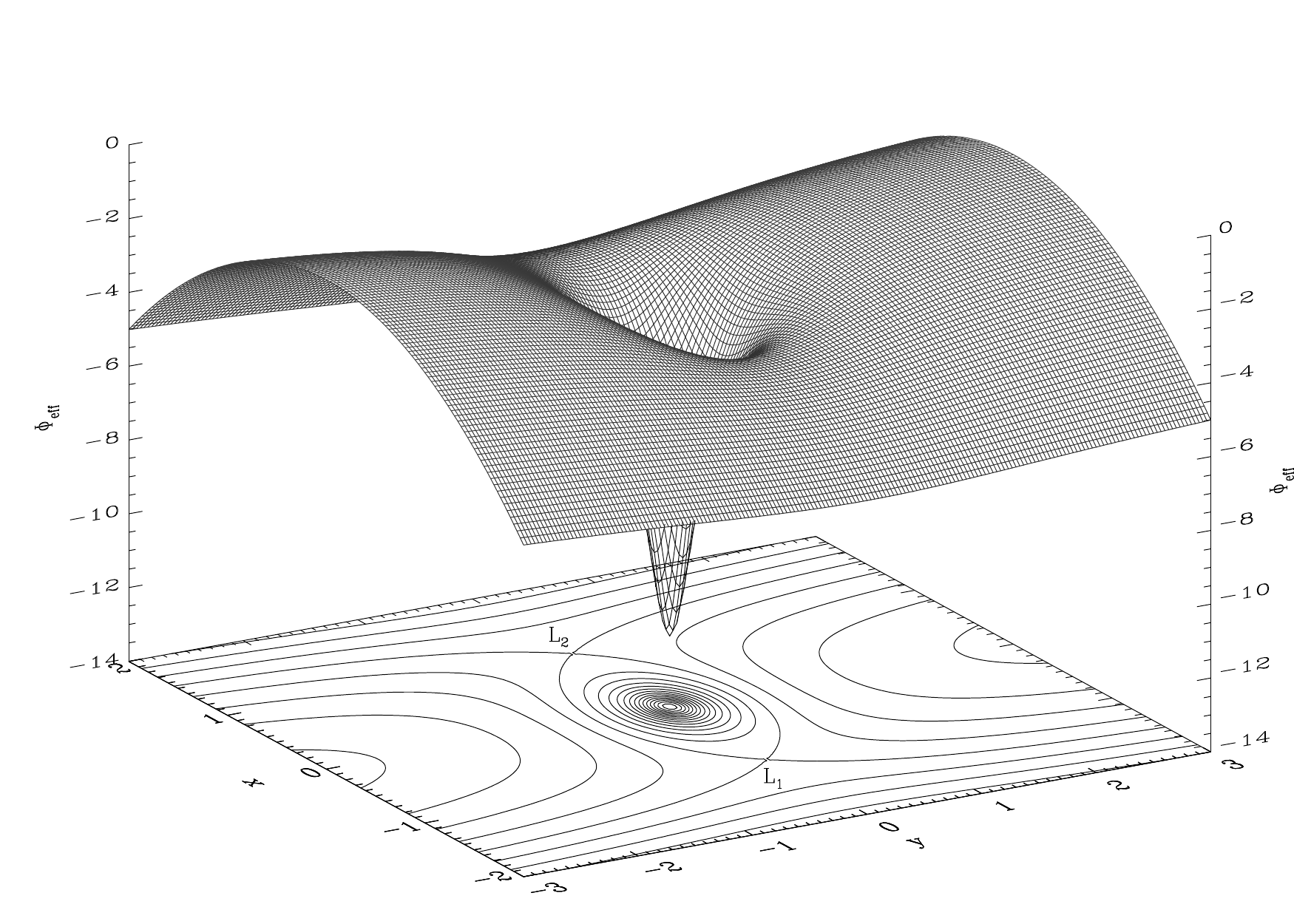} 
\caption{Effective potential in the tidal approximation ($z=0$ plane). The Lagrangian points 
at $L_1=(-1,0)$ and $L_2=(1,0)$ can be seen. The escapers pass these saddle points while
they leak out. The equipotential lines connecting them mark the tidal boundary of the star cluster. 
The details of the model are given in Section 3.} 
\label{fig:effpot}
\end{figure*}

\begin{equation}
\Phi_{\rm eff}(x,y,z) = \Phi_{\mathrm{cl}}(x,y,z) + \frac{1}{2}(\kappa^2 - 4\omega^2) x^2 
+ \frac{1}{2}\nu^2 z^2 \label{eq:effpot}
\end{equation}

\noindent
is the effective potential, which is plotted in Figure 1 for the 2D case. Other isolating integrals are not given in the form of a simple analytical expression. However, some solutions of 
(\ref{eq:eqm1})  - (\ref{eq:eqm3}) are subject to a third 
integral (H\'enon \& Heiles 1964), as has been demonstrated
numerically by Fukushige \& Heggie (2000), who calculated a Poincar\'e surface of section.
In principle, one may obtain power series expansions of such third integrals, see, e.g.
the original works by Gustavson (1966) and Finkler, Jones \& Sowell (1990) and the
review in Moser (1968). 
Also, third integrals can be related to the existence of Killing tensor fields which are well-known 
in General Relativity (Clementi \& Pettini 2002).
At last, the tidal radius (King 1962) 

\begin{equation}
r_t=\left(\frac{GM_{cl}}{4\omega^2 - \kappa^2}\right)^{1/3} \label{eq:rtidal}
\end{equation}

\noindent
where $M_{cl}$ is the star cluster mass, provides a fundamental length scale of the problem.
It is the distance from the origin of coordinates to the Lagrangian points $L_1$ and 
$L_2$ (which lie on the $x$ axis, see Figure \ref{fig:effpot}).

%These can be 
%derived from an effective potential in the epicyclic approximation:

%\begin{equation}
%\Phi_\mathrm{eff}(x,y,z)=\Phi_{\mathrm{cl}}(x,y,z) + \frac{1}{2}\mu^2 x^2 
%+ \frac{1}{2}\nu^2 z^2 + \ \mathcal{O}(xz^2)  \label{eq:phieff} + \mathrm{const}  
%\end{equation}

%\noindent
%where the coordinates are relative to the star cluster centre and
%$\Phi_{\mathrm{cl}}$ is the star cluster potential. Since the expansion is about the
%minimum of $\Phi_{\rm eff}$, where the first partial derivatives vanish,
%there are no first-degree terms in the Taylor expansion. The term $\propto xz$
%vanishes because $\Phi_{\rm eff}$ is symmetric in $z$. The term $\propto x^2$
%arises from a combination of centrifugal and tidal forces. For an illustration
%using a Plummer potential for the star cluster, see
%Figure \ref{fig:equipot}, which also shows the location of the effective potential's 
%nearest equilibrium points with respect to the star cluster centre, which are
%covered by the approximation: The Lagrangian points $L_1$ and $L_2$, where 
%$\nabla \Phi_{\rm eff} = 0$.
%They are saddle points of the effective potential, i.e. the Hessian is neither positive
%nor negative definite.

\section{The model}

The characteristic frequencies $\omega, \kappa$ and $\nu$ arise as properties of the
galactic gravitational potential. Throughout the paper, we use the values of the 
characteristic frequencies in the solar neighborhood. 
All of them can be expressed in terms of Oort's constants $A$ and $B$ (see, e.g., Binney \& Tremaine
1987): $\omega^2 = (A-B)^2$, 
$\kappa^2 = -4B(A-B)$, $\kappa^2 - 4\omega^2 = -4A(A-B)$, $\nu^2 = 4\pi G\rho_g+2(A^2-B^2)$. 
The vertical frequency $\nu$ can be derived from the Poisson equation for an axisymmetric system (see Oort 1965) and $\rho_g$ is the local Galactic density, which contributes to the
dominant first term. We obtain both dimensionless parameters $\kappa^2/\omega^2\simeq 1.8$ 
and $\nu^2/\omega^2\simeq 7.6$ using the values of Oort's constants given in 
Feast \& Whitelock (1997) and the value for local Galactic density given in 
Holmberg \& Flynn (2000). It is then convenient to choose the following
system of units:

\begin{equation}
G = 1, \ \ \ \ \ \omega = 1, \ \ \ \ \ M_{cl} = 2.2
\end{equation}

\noindent
The resulting length unit is the tidal radius $r_t$ and the formulation of the dynamical problem 
with its equations of motion is completely dimensionless. For the star cluster, we use 
a Plummer model, the most simple analytic model for a star cluster (see appendix A). 
The density profiles of King models which have a cutoff radius, where the
density drops to zero, fit the measured density profiles of globular clusters better than Plummer 
models (King 1966). Since they are tidally limited by construction, they are at first glance ideally
suited for our purpose. However, the gravitational force field can only be tabulated from a
numerical integration of a non-linear differential equation. We have made a compromise which is
not relevant for the interesting physics: We choose the Plummer radius in such a way, that
the Plummer model (see Appendix A) is the best fit to a King model with $W_0=4$ 
(i.e. with concentration $c = \log_{10}(r_t/r_K) \simeq 0.840$), which completely fills the Roche
lobe in the tidal field, i.e. the density of the King model approaches zero at the tidal radius
(\ref{eq:rtidal}). The fit of the density profiles is quite acceptable 
for density contrasts of $\log_{10}(\rho_c/\rho(r))\lesssim 3$, where $\rho_c$ and $\rho(r)$ are the 
central density and the density as a function of radius, respectively. 
For the deviation between the King density profile and the Plummer fit we obtain
$(\rho_{Pl}(r) - \rho_K(r))/\rho_c < 1.2 \%$.
The ratio of the Plummer radius to the King radius and the ``concentration'' of the Plummer model 
(which can only be defined because of the existence of a tidal radius) are then 
$r_{Pl}/r_K \simeq 1.257$ and $c_{Pl}=\log_{10}(r_t/r_{Pl}) \simeq 0.741$, respectively. In our units, the Plummer radius is therefore $r_{Pl} \simeq 0.182$. 

As a more technical remark, we note that we used an $8$th-order Runge-Kutta scheme for the 
orbit integrations. The relative error in the Jacobian $C$ was always limited to 
$\Delta C/C < 10^{-12}$ for all orbit integrations.

\section{Poincar\'e surfaces of section and the Coriolis asymmetry}

\begin{figure*}
\includegraphics[width=0.9\textwidth]{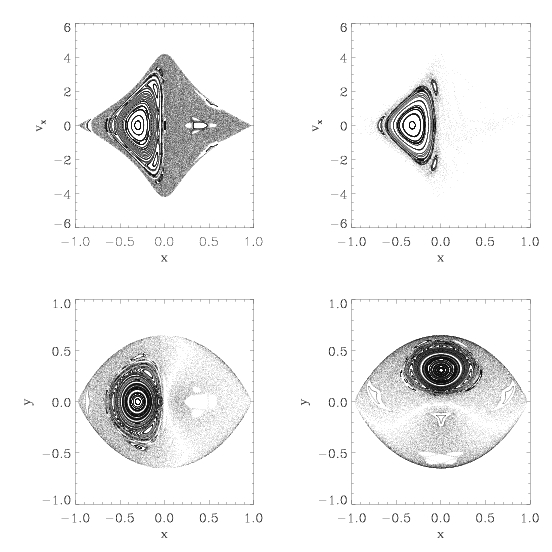} 
\caption{Poincar\'e surfaces of section. 
Top left: At $\widehat{C} = 0$ for orbits crossing $y=0$ with $\dot{y}>0$, 
Top right: Same as top left, but at $\widehat{C} = 0.1$, 
Bottom left: At $\widehat{C} = 0$ for orbits crossing $\dot{x}=0$ with $\dot{y}>0$, 
Bottom right: At $\widehat{C} = 0$ for orbits crossing
$\dot{y}=0$ with $\dot{x}>0$. The variable $\widehat{C}$ is defined in Equation (\ref{eq:chat}).} 
\label{fig:poincare}
\end{figure*}

\begin{figure}
\includegraphics[width=0.45\textwidth]{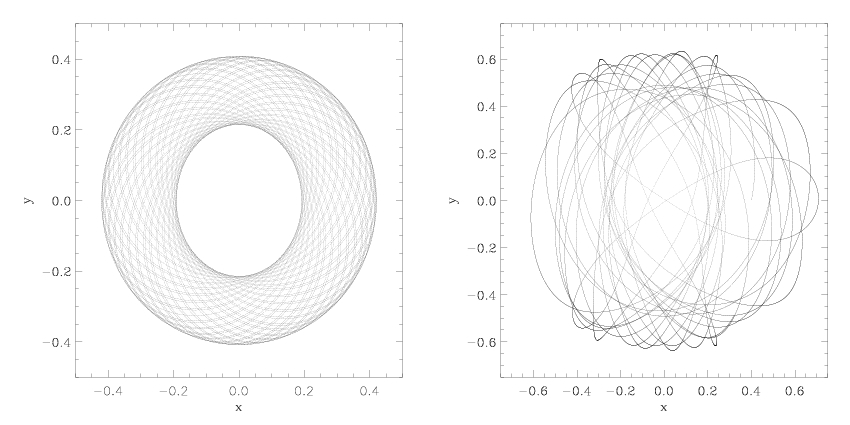} 
\caption{The two main types of orbits at $\widehat{C} = 0$. 
Left: Regular retrograde orbit, Right: Chaotic prograde orbit. The variable $\widehat{C}$ is 
defined in Equation (\ref{eq:chat}).}
\label{fig:orbit2d}
\end{figure}
% Regular from $(-0.4, 0.0, 0.4)$
% Chaotic from $(0.4, 0.0, 0.4)$

\begin{figure}
\includegraphics[width=0.5\textwidth]{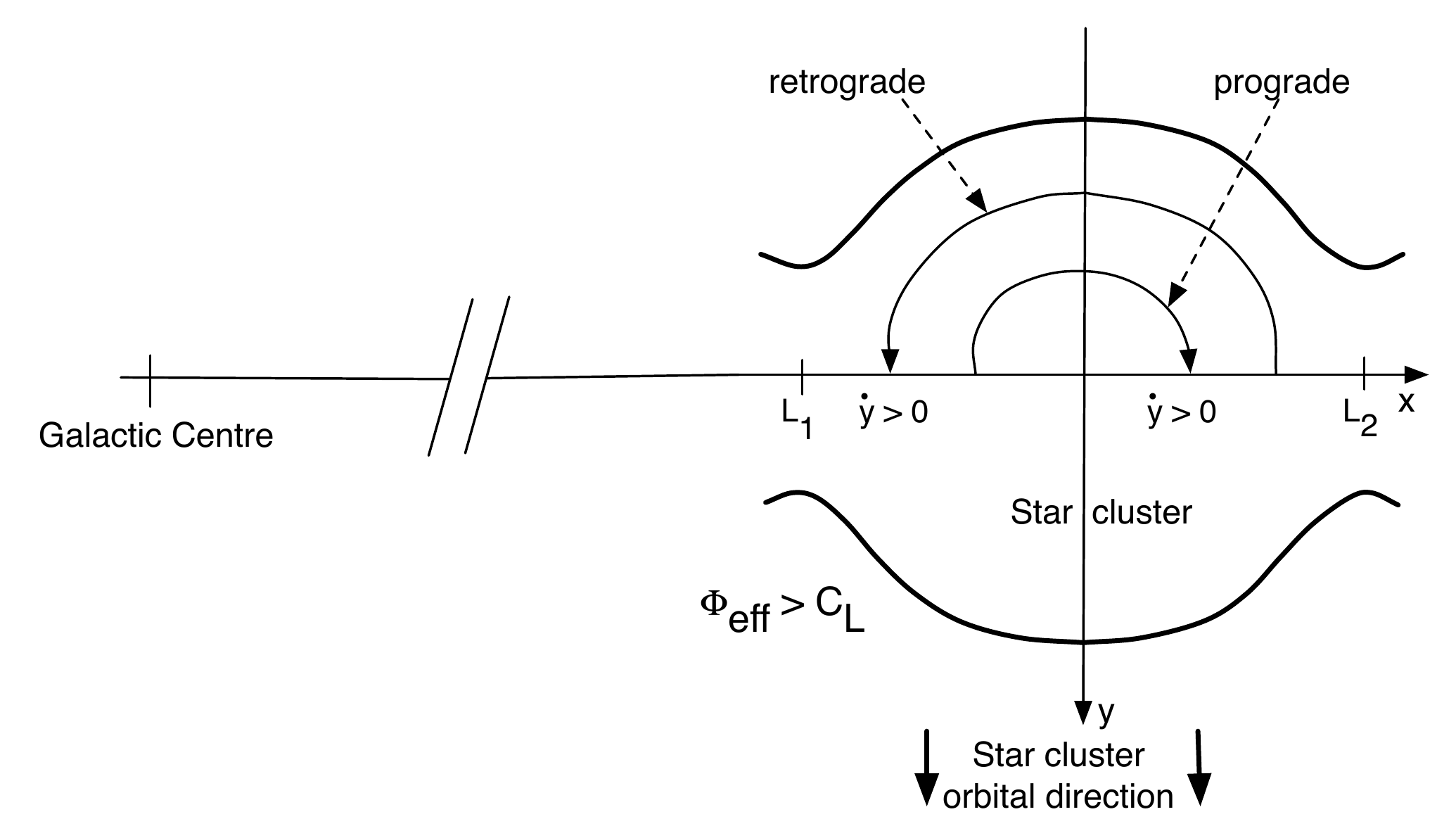} 
\caption{Sketch of the coordinate system. The escapers leak out through the openings in
the equipotential surfaces passing either $L_1$ or $L_2$. Only schematically, two orbits are
shown which cross the $x$ axis with $\dot{y}>0$.} 
\label{fig:sketch}
\end{figure}

A critical Jacobi constant 

\begin{equation}
C_L = \Phi_{\rm eff}(r_t,0,0)=\Phi_{\rm eff}(-r_t,0,0) \label{eq:CL}
\end{equation}

\noindent
is given by the value of 
the effective potential (\ref{eq:effpot}) at the 
Lagrangian points $L_1$ and $L_2$. For our model we have $C_L = -3.264444506$.
For a Jacobian $C > C_L$ the equipotential surfaces are open and particles can escape.
Furthermore, we define the dimensionless deviation from $C_L$ by

\begin{equation}
\widehat{C} = (C_L - C)/C_L, \label{eq:chat}
\end{equation}

\noindent
where $C$ is some other value of the Jacobian. The dimensionless deviation $\widehat{C}$ 
is positive for $C > C_L$ if $C$ and $C_L$ are both negative, which is always the case in 
this paper.
A first insight can be gained by calculating Poincar\'e surfaces of section which are
shown in Figure \ref{fig:poincare} for two different Jacobi constants: The upper left surface of 
section is at the critical Jacobian $C_L$ at which all orbits still remain within a bounded
area in phase space and we have no escapers. We can see that this is a system with 
divided phase space, i.e. we have both chaotic and regular orbits. It is striking that
the left half of the surface of section is almost completely occupied by regular, 
quasiperiodic orbits. These orbits are retrograde with respect to the orbit of the star cluster around
the galactic centre (Fukushige \& Heggie 2000), as can be seen by looking at the sketch in 
Figure \ref{fig:sketch}, and they  are subject to a third integral of motion. On the other hand, 
most of the prograde orbits are chaotic apart from a few smaller regular islands. 
Since the effective potential is mirror-symmetrical with respect to both the $x$- and $y$-axes 
(see Figure \ref{fig:effpot}),  the asymmetry seen in the surfaces of section must be due to 
the Coriolis forces. 
Thus such a behaviour might be termed a ``Coriolis asymmetry'' (cf. Innanen 1980). 
The Coriolis forces are special in the sense that their direction is not perpendicular to the 
tangent plane to the equipotential surfaces but to the velocity of a particle.
The upper right surface of section is at a higher Jacobi constant. The particles can
leak out through the openings in the equipotential surfaces and escape towards
infinity (positive $x$-direction) or the galactic centre (negative $x$-direction). 
It is remarkable, that only the chaotic orbits escape, while the regular, quasiperiodic
orbits remain within the tidal boundary of the star cluster, since the third integral
restricts their accessible phase space and hinders their escape. In star clusters, two-body 
relaxation may scatter stellar orbits beyond the critical Jacobi constant. However,
if the orbits are retrograde, the stars will remain bound to the star cluster with high probability 
until two-body relaxation further scatters them into the escaping phase space. 
The two lower surfaces of section in Figure \ref{fig:poincare} indicate the orbital structure in
position space at $C=C_L$, where all orbits are restricted to the region within the
almond-shaped tidal boundary of the star cluster (cf. Figure \ref{fig:effpot}). 
One notes that certain parts of the chaotic regions of the lower surfaces of section
are less densely filled by stellar orbits which is a common property of dynamical 
systems.

Figure \ref{fig:orbit2d} shows typical examples of the two main types of orbits. 
The regular orbit resembles a rosette orbit in an axisymmetric potential or a loop orbit which
suggests that the third integral is some sort of generalization of angular momentum
(Binney \& Tremaine 1987). We found indeed that the angular momentum is approximately
conserved for the left orbit of Figure \ref{fig:orbit2d}, while it is not at all conserved for the right
orbit. Other types of orbits can be found which are associated 
with the smaller regular islands in the surfaces of section and which may look rather interesting.
At last, we remark that the difference between retrograde and prograde orbits appears
prominently in $N$-body models of dissolving {\it rotating} star clusters within the 
framework of the tidal approximation: Where either the regular or the chaotic domains in
phase space are more strongly occupied by stellar orbits due to the existence of
a net angular momentum of the cluster in one direction (see Ernst et al. 2007).

\section{The basins of escape}

\begin{figure*}
\includegraphics[width=0.85\textwidth]{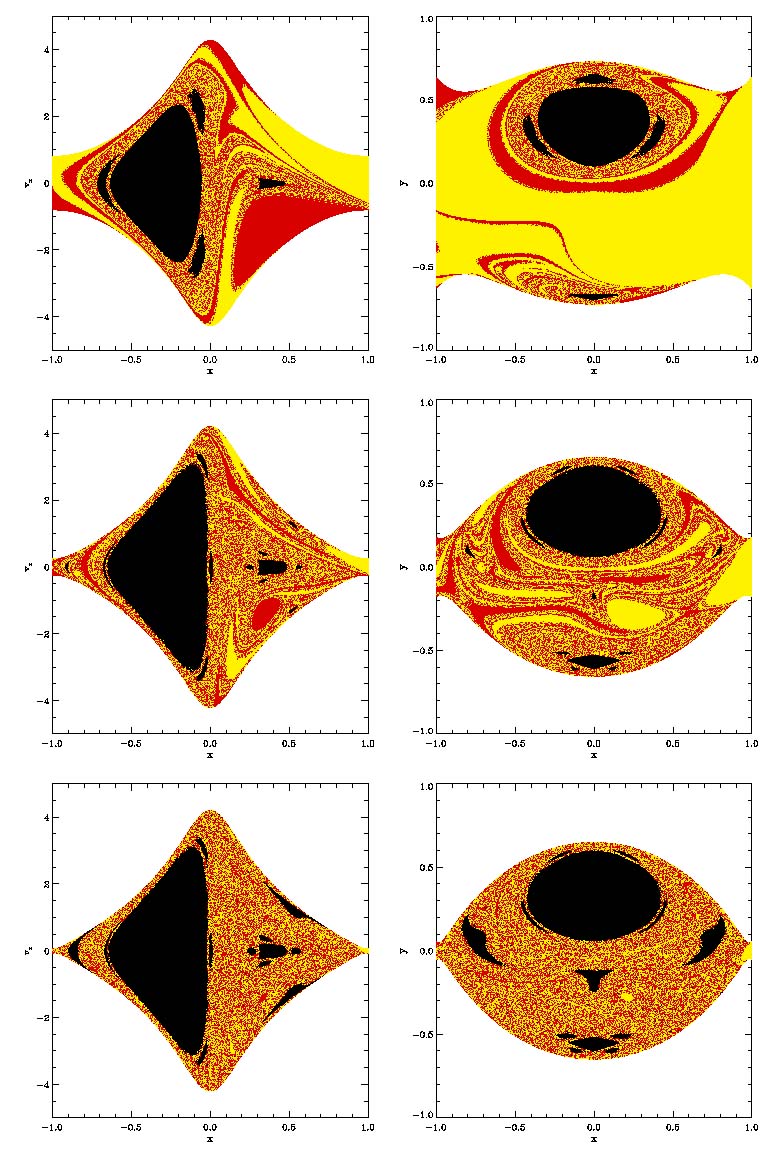} 
\caption{The basins of escape. 
Top left: At $\widehat{C} = 0.1$ for orbits crossing $y = 0$ with $\dot{y}>0$,
Top right: At $\widehat{C} = 0.1$ for orbits crossing $\dot{y} = 0$ with $\dot{x}>0$, 
Middle row: As the upper row, but at $\widehat{C} = 0.01$, 
Botton row: As the middle row, but at $\widehat{C} = 0.001$.
The red regions denote initial conditions, where the escaping stars pass $L_1$ while the yellow 
regions represent initial conditions, where the escaping stars pass $L_2$. The black regions
show those initial conditions, where stars do not escape.
The variable $\widehat{C}$ is defined in Equation (\ref{eq:chat}).}
\label{fig:basins}
\end{figure*}

Figure \ref{fig:basins} shows the basins of escape for a tidally limited star cluster within the 
framework of the
tidal approximation. For the plots on the left-hand side, $\sim 3\times 10^5$ orbits have been
integrated; for the plots on the right-hand side, it were $\sim 6 - 8 \times 10^5$ orbits,
depending on the area of the surface of section containing initial conditions. 
The phase space is divided into the escaping phase space (red and yellow regions) and 
the non-escaping phase space (black regions).
The red regions denote initial conditions, where the escaping stars pass $L_1$ while the yellow 
regions represent initial conditions, where the escaping stars pass $L_2$. The black regions
show those initial conditions, where stars do not escape.
These are first and foremost the regular regions where a third integral is present. Note that the 
stable manifold of the chaotic saddle is not marked in black, since it is of Lebesgue measure
zero (cf. Section 6).
Since there exist regular regions (with KAM tori), the system is non-hyperbolic, i.e. there exist 
stable periodic orbits with corresponding
elliptic points in the surfaces of section (cf. Section 4 and Figure \ref{fig:poincare}). 
We can see that in the escaping phase space there exist regions, where we have a very 
sensitive dependance of the escape process on the initial conditions, i.e. a slight change of 
the initial conditions makes the star
escape through the opposite Lagrangian point. This is the classical indication of chaos.
It is interesting to note that these regions arise from immediate vicinity of the
black regions where orbits are regular.
In these domains of phase space the red and yellow regions are completely intertwined
with respect to each other: The boundary between these regions is fractal.
The volume in phase space occupied by these
regions (with sensitive dependance on the initial conditions) increases as the 
Jacobian $C$ approaches the critical Jacobi constant $C_L$ (i.e. in the limit $\widehat{C} \rightarrow 0$) and the exits become smaller. At $\widehat{C} \rightarrow 0$ there is a maximal
``fractalization'' of phase space; when the system approaches the limit of small exits the 
basins become uncertain (Aguirre \& Sanju\'an 2003). 
The term ``uncertain'' means that we become unable to follow the real trajectory of a particle by means of numerical integration.\footnote{In other words, the computer fails here to be a Laplacian demon 
(Laplace 1814).}
Moreover, the following theorem can be formulated: ``{\it For all points $P$ in the escaping phase space 
of an open Hamiltonian system, 
and for all $\delta>0$ (precision of the experiment), there exists a critical size of the exits $w_c>0$ 
such that for all $w \leq w_c$ we can find a point $P'$ in a ball centreed in P and radius $\delta$
that belongs to a different basin than $P$}'' (Aguirre \& Sanju\'an 2003).
 
 \begin{table}
\begin{center}
\begin{tabular}{llll}
\hline
Plot (Figure \ref{fig:basins}) & Black & Red & Yellow  \\
\hline
Top left & 25.4 & 36.6 & 38.0 \\
Top right & 12.5 & 19.3 & 68.2 \\
Middle left & 36.9 & 31.5 & 31.5 \\
Middle right & 21.4 & 36.8 & 41.8 \\
Bottom left & 40.9 & 29.5 & 29.6 \\
Bottom right & 27.2 & 36.2 & 36.6 \\
\hline
\end{tabular} 
\end{center}
\caption{Fraction of orbits (in percent) belonging to the intersection of the basins of escape 
with Poincar\'e surfaces of section which are shown in Figure \ref{fig:basins}.}
\label{tab:fraction}
\end{table}

 \begin{figure}
\includegraphics[width=0.45\textwidth]{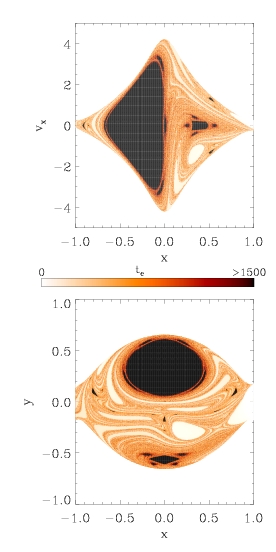} 
\caption{Distribution of escape times $t_e$ on surfaces of section for $\widehat{C}=0.01$.
Top: For the $x-v_x$ surface of section of Figure \ref{fig:basins}, Bottom: For the $x-y$ surface of section of Figure \ref{fig:basins}. The darker the color, the longer the escape time.}
\label{fig:times}
\end{figure}

 \begin{figure}
\includegraphics[width=0.45\textwidth]{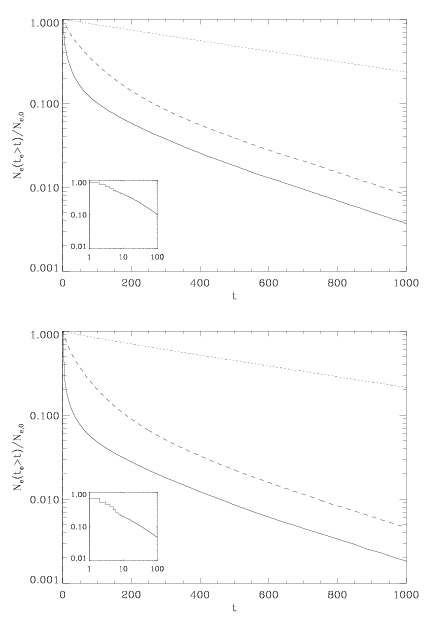} 
\caption{Histogram of the fraction of remaining (non-escaped) orbits $N_e(t_e>t)/N_{e,0}$ after time $t$. 
Top: For the $x-v_x$ surfaces of section of Figure \ref{fig:times}, Bottom: For the $x-y$ surfaces of section of Figure \ref{fig:times},
Solid:  $\widehat{C}=0.1$, Dashed:  $\widehat{C}=0.01$, Dotted:  $\widehat{C}= 0.001$.
The inlays with two logarithmic axes show the early phase for the solid line 
(i.e. for $\widehat{C}=0.1$). The variable $\widehat{C}$ is defined in Equation (\ref{eq:chat}).}
\label{fig:tcum3}
\end{figure}

Table \ref{tab:fraction} shows the fraction of orbits (in percent) belonging to the intersection
of the basins of escape with Poincar\'e surfaces of section which are shown in Figure \ref{fig:basins}. 
It can be seen that in the limit
$\widehat{C}\rightarrow 0$ the fractions of particles passing $L_1$ and $L_2$
tend to be equal while this must not be the case if there are large areas without
sensitive dependance of the escape process on the initial conditions.

Figure \ref{fig:times} shows how the escape times are distributed on surfaces of section.
The longest escape times correspond to initial conditions near the boundaries between the 
basins of escape of Figure \ref{fig:basins}. The shortest escape times have been measured for the 
ordered regions without sensitive dependance on the initial conditions, i.e. those
far away from the fractal basin boundaries.

Figure \ref{fig:tcum3} shows the fraction of remaining (non-escaped) orbits $N_e(t_e>t)/N_{e,0}$
after time $t$ corresponding to the basins of escape shown in Figure \ref{fig:basins}. Only the
escaping orbits have been used for the statistics. 
The orbits corresponding to the regions without sensitive dependance on the initial conditions
shown in Figures \ref{fig:basins} have short escape times as can be seen in Figure \ref{fig:times}.
For these orbits, the decay law is a power law as shown in the inlays for the solid 
curve ($\widehat{C}=0.1$). 
On the other hand, the decay law is exponential for the chaotic orbits near the fractal basin boundaries (i.e. the orbits with long escape times which correspond to the regions with 
sensitive dependance on the initial conditions). The slopes of the exponentials (i.e. the 
decay constants) depend on the value of $\widehat{C}$ but are identical for both surfaces of section.
The exponential decay law indicates that the underlying process is of a statistical nature similar to the radioactive decay of unstable nuclides or that of bubbles in beer foam. 

\section{The chaotic saddle and its invariant manifolds}

\begin{figure*}
\includegraphics[width=0.85\textwidth]{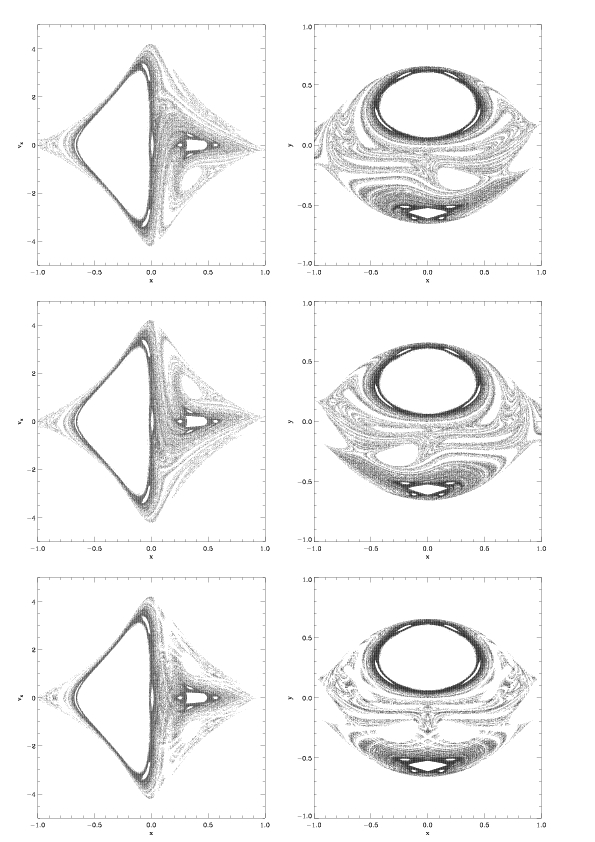} 
\caption{The non-hyperbolic invariant set and its stable and unstable manifolds at 
$\widehat{C} = 0.01$. 
Top row: Stable manifold,
Middle row: Unstable manifold,
Bottom row: Invariant set. 
The variable $\widehat{C}$ is defined in Equation (\ref{eq:chat}).}
\label{fig:saddle}
\end{figure*}

The stable manifold of the chaotic saddle is shown in the top row of Figure \ref{fig:saddle} for 
two Poincar\'e surfaces of section.
The stable manifold coincides with the fractal basin boundaries of Figure \ref{fig:basins}
and therefore acts as a separatrix between the exit basins..
With data points of finite size, the top row shows orbits which do not escape for time 
$t \rightarrow \infty$, although, strictly speaking, their Lebesgue measure is zero.
The unstable manifold of the chaotic saddle is shown in the middle row of 
Figure \ref{fig:saddle}. These are orbits which do not escape for time $t \rightarrow -\infty$. 
Note that the stable and unstable manifolds are symmetric with respect to each other, 
since the equations of motion (\ref{eq:eqm1}) - (\ref{eq:eqm3}) are time-symmetric.
For the plots in the middle row of Figure \ref{fig:saddle}, the sign of the time step in the
Runge-Kutta integrator has been reversed,
The bottom row of Figure \ref{fig:saddle} shows the intersection of the chaotic saddle 
(i.e. a non-hyperbolic chaotic invariant set) with the two Poincar\'e surfaces of section. 
It is the invariant set of
non-escaping orbits for time $t \rightarrow \infty$ and $t \rightarrow -\infty$.
The chaotic saddle has the form of a Cantor set (Cantor 1884) which is formed by the 
intersection of its stable 
and unstable manifolds. The fact that the system is non-hyperbolic implies that there are 
tangencies between the stable and unstable manifolds, i.e. that their angle is not always 
bounded away from zero (Lai et al. 1993). The unstable (hyperbolic) points in the
intersection of the Poincar\'e surface of section with the chaotic saddle correspond to 
unstable periodic orbits. As is well-known (e.g. Contopoulos 2002),
these introduce chaos into the system since they repel the orbits
in their neighbourhood in the direction of their unstable eigenvectors.
The remarkable similarity of our system with the H\'enon-Heiles system (see Aguirre, Vallejo \&
Sanju\'an 2001)
is that the fractal dimension of the chaotic saddle tends to three (i.e. the black areas in Figure 
\ref{fig:saddle} grow until we have a maximal fractalization of the phase space, cf. Aguirre 
\& Sanjuan 2003), which is the dimension of the hypersurface of phase space with 
constant Jacobian.
At $\widehat{C}=0$ there is a sudden transition where the non-hyperbolic
invariant set abruptly fills the whole non-regular subset of phase space within the last 
closed equipotential surface and no escape is possible any more. This situation
can be seen in the Poincar\'e surfaces of section in Figure \ref{fig:poincare} in
Section 4 which shows the chaotic domains of phase space as a dotted area.

\section{Discussion and conclusions}

We studied the chaotic dynamics within a star cluster which is embedded
in the tidal field of a galaxy.
We calculated within the framework of the tidal approximation Poincar\'e surfaces
of section, the basins of escape, the chaotic saddle and its stable and unstable
manifolds as well. The system is non-hyperbolic which has important consequences
for the dynamics, i.e. there are orbits which do not escape if relaxation is neglected.
These are mainly the retrograde orbits as has been shown earlier by
Fukushige \& Heggie (2000) and, more recently, in the $N$-body study by Ernst et al. (2007). 
The escape times are longest for initial conditions near 
the fractal basin boundaries. The decay law is a power law for those stars which escape
from the regions without sensitive dependance on the initial conditions in Figure \ref{fig:basins}
(i.e. with short escape times, as can be seen in Figure \ref{fig:times}).
On the other hand, the decay law is exponential for orbits which escape from the regions with
sensitive dependance on the initial conditions (i.e. with long escape times).
The effect of relaxation (i.e. a diffusion in the Jacobian and the third integral among
different stellar orbits) on the chaotic dynamics 
which we investigated in this work may be a very interesting topic for future research.

\section{Acknowledgements}

We thank Dr. Juan Carlos Vallejo and Prof. Burkhard Fuchs for pointing out some references
to us, which were necessary for this work. Also, we thank Prof. Rudolf Dvorak for his comments
on the manuscript. AE would like to thank his father for a discussion 
and Dr. Patrick Glaschke for many inspiring discussions. Furthermore, he gratefully acknowledges 
support by the International Max Planck Research School (IMPRS) for Astronomy and Cosmic Physics 
at the University of Heidelberg.

\appendix

\section{The isolated Plummer model}

The most well-known self-consistent model of a spherically symmetric 
star cluster is the Plummer model (Plummer 1911). It is given by

\begin{eqnarray}
\Phi(y) &=& -\Phi_0\frac{1}{\sqrt{1+y^2}} = -\Phi_0\cos\phi \label{eq:plummerpotential} \\
M(y) &=& M\frac{y^3}{\left(1+y^2\right)^{3/2}} = M\sin^3\phi \label{eq:plummermass} \\
\rho(y) &=& \rho_0 \frac{1}{\left(1+y^2\right)^{5/2}} =\rho_0\cos^5\phi \label{eq:plummerdensity}
\end{eqnarray}

\noindent
with $y=r/r_{{\rm Pl}}\geq 0$, $\phi=\arctan(y)$ and the Plummer radius
%$y=\tan\phi$

\begin{equation}
r_{{\rm Pl}} = \frac{GM}{\Phi_0} = \left( \frac{3M}{4\pi\rho_0}\right)^{1/3}
\end{equation}

\noindent
where $M$ is the total mass, $-\Phi_0$ is the central potential and $\rho_0$ the central 
density, all of them being finite. From (\ref{eq:plummermass}) we find the
half mass radius

\begin{equation}
y_h 
%= \left(2^{2/3} - 1\right)^{-1/2} 
%= \tan\left(\arcsin\sqrt[3]{1/2}\right)\simeq 1.30476603
= \tan\left[\arcsin\left(2^{-1/3}\right)\right]\simeq 1.30476603
\end{equation}

\noindent
The velocity dispersion can be obtained by integrating the Jeans equation of 
hydrostatic equilibrium. It is given by

\begin{equation}
\sigma^2(y) = \frac{\Phi_0}{6} \frac{1}{ \sqrt{1 + y^2}} = \frac{\Phi_0}{6} \cos\phi
\end{equation}

\noindent
Therefore the relation

\begin{equation}
v_e^2(y) = 12\sigma^2(y),
\end{equation}

\noindent
where $v_e=\sqrt{2\vert\Phi(y)\vert}$ is the escape speed from an isolated Plummer model,
strictly holds at any radius (cf. Section 1 of this paper). One can show that the Plummer model is the 
only model which has this property. 

\bsp

\label{lastpage}


\begin{thebibliography}{99}
\hypertarget{mybib}{}
\pdfbookmark[0]{Bibliography}{mybib} 

\bibitem{} Ambartsumian V. A., 1938, Ann. Leningrad State Univ., 22, 19
\bibitem{} Aguirre J., Vallejo J. C., Sanju\'an M. A. F., 2001, Phys. Rev. E, 64, 066208
\bibitem{} Aguirre J., Sanju\'an M. A. F., 2003, Phys. Rev. E, 67, 056201
\bibitem{} Aguirre J., Vallejo J. C., Sanju\'an M. A. F., 2003, Int. J. Mod. Phys. B, 17, 4171 
\bibitem{} Aguirre J., 2004, PhD thesis, Universidad Rey Juan Carlos, Spain
\bibitem{} Baumgardt H., 2001, MNRAS, 325, 1323
\bibitem{} Benet L., Trautmann D., Seligman T. H., 1997, Cel. Mech. Dyn. Astron., 66, 203 
\bibitem{} Benet L., Seligman T. H., Trautmann D., 1999, Cel. Mech. Dyn. Astron., 71, 167 
\bibitem{} Binney J., Tremaine S., 1987, {\it Galactic Dynamics}, Princeton Univ. Press 
\bibitem{} Bleher S., Grebogi C., Ott, E., Brown R., 1988, Phys. Rev. A, 38, 930 
\bibitem{} Bleher S., Ott E., Grebogi C., 1989, Phys. Rev. Lett., 63, 919 
%\bibitem{} Bleher S. Grebogi C., Ott E., 1990, Physica D, 46, 87 
\bibitem{} Cantor G., 1884, Acta Mathematica, 4, 381. English transl. in
ed. Edgar G. A., 1993,  {\it Classics on Fractals}, Addison-Wesley
\bibitem{} Chandrasekhar S., 1942, {\it Principles of Stellar Dynamics}, Univ. Chicago
Press, Chicago
\bibitem{} Churchill R. C., Pecelli G., Rod D. L., 1975, J. Diff. Eq., 17, 329 
\bibitem{} Clementi C., Pettini M., 2002, Cel. Mech. Dyn. Astron., 84, 263
\bibitem{} Contopoulos G., 1990, Astron. Astrophys., 231, 41 
\bibitem{} Contopoulos G., Kaufmann D., 1992, Astron. Astrophys., 253, 379 
\bibitem{} Contopoulos G., 2002, {\it Order and Chaos in Dynamical Astronomy}, Berlin (Springer)
\bibitem{} De Moura A. P. S., Letelier P. S., 1999, Phys. Lett. A, 256, 362 
\bibitem{} Eckhardt B., Jung C., 1986, J. Phys. A: Math. Gen., 19, L829 
\bibitem{} Eckhardt B., 1987, J. Phys. A: Math. Gen., 20, 5971 
\bibitem{} Ernst A., Glaschke P., Fiestas J., Just A., Spurzem R., 2007, MNRAS, 377, 465 
\bibitem{} Feast M., Whitelock P., 1997, MNRAS, 291, 683 
\bibitem{} Finkler P., Jones C. E., Sowell G. A., 1990, Phys. Rev. A, 42, 1931
\bibitem{} Fukushige T., Heggie D. C., 2000, MNRAS, 318, 753 
\bibitem{} Glaschke P., 2006, PhD thesis, University of Heidelberg, 
http://www.ub.uni-heidelberg.de/archiv/6553
\bibitem{} Gustavson F., 1966, AJ, 71, 670 
\bibitem{} H\'enon M., 1960, Ann. Astrophys., 23, 668
\bibitem{} H\'enon M, Heiles C., 1964, AJ, 69, 73
\bibitem{} H\'enon M., 1969, A\&A, 2, 151
\bibitem{} Holmberg J., Flynn C., 2000, MNRAS, 313,209 
\bibitem{} Hunt B. R., Ott E., Yorke J. A., 1996, Phys. Rev. E, 54, 4819
\bibitem{} Innanen K. A., 1980, Ap. J., 85, 81 
\bibitem{} Jung. C., 1987, J. Phys. A: Math. Gen., 20, 1719 
\bibitem{} Jung C., Scholz H. J., 1987, J. Phys. A: Math. Gen., 20, 3607 
\bibitem{} Jung C., Pott S., 1989, J. Phys. A: Math. Gen., 22, 2925 
\bibitem{} Jung C., Ziemniak E., 1992, J. Phys. A: Math. Gen., 25, 3929 
\bibitem{} King I. R., 1959, AJ, 64, 351 
\bibitem{} King I. R., 1962, AJ, 67, 471 
\bibitem{} King I. R., 1966, AJ, 71, 64 
\bibitem{} Lai Y.-C., Grebogi C., Yorke J. A., Kan I., 1993, Nonlinearity, 6, 779 
\bibitem{} Lai Y.-C., 1997, Phys. Rev. E, 56, 6531
\bibitem{} Laplace, 1814, Th\'eorie analytique des probabilit\'es. In: {\it Oeuvres Compl\'etes de
Laplace}, Volume VII., Paris, 1820 (Gauthier-Villars)
\bibitem{} Moser J. K., 1968, {\it Lectures on Hamiltonian Systems}, Mem. AMS, 81, 1. Reprinted in
eds. MacKay R. S. and Meiss J. D., 1987, {\it Hamiltonian Dynamical Systems}, Bristol (Adam Hilger)
\bibitem{} Motter A. E. Lai Y., 2001, Phys. Rev. E, 65, 015205 
\bibitem{} Navarro J. F., Henrard J., 2001, A\&A, 369, 1112 
\bibitem{} Oort J. H., 1965, \emph{Stellar Dynamics},
in: \emph{Galactic Structure}, eds. A. Blaauw \& M. Schmidt, 
Univ. Chicago Press, p. 455
\bibitem{} Ott E., Sommerer J. C., Alexander J. C. Kan I., Yorke J. A., 1993,
Phys. Rev. Lett., 71, 4134 
\bibitem{} Plummer H. C., 1911, MNRAS, 71, 460
\bibitem{} Rod D. L., 1973, J. Diff. Eq., 14, 129 
\bibitem{} Schneider J., T\'el T., Neufeld Z., 2002, Phys. Rev. E, 66, 066218 
\bibitem{} Seoane Sep\'ulveda J. M., 2007, PhD thesis, Universidad Rey Juan Carlos, Spain
\bibitem{} Siegel C. L., Moser J. K., 1971, {\it Lectures on Celestial Mechanics}, Springer Edition,
Berlin, Heidelberg, New York
\bibitem{} Siopis C., Kandrup H. E., Contopoulos G., 1997, Cel. Mech. Dyn. Astron., 65, 57 
\bibitem{} Sommerer J. C., Ott E., 1993, Nature, 365, 138 
\bibitem{} Sommerer J. C., Ott E., 1996, Phys. Lett. A, 214, 243 
\bibitem{} Spitzer L. Jr., Shapiro S. L., 1972, Ap. J., 173, 529 
\bibitem{} Spurzem R., Giersz M., Takahashi K., Ernst A., 2005, MNRAS, 364, 948
\bibitem{} Stumpff K., 1965, {\it Himmelsmechanik}, Vol. 2, VEB Deutscher Verlag der
Wissenschaften, Berlin
\bibitem{} Szebehely V., 1967, {\it Theory of orbits. The restricted problem of three bodies},
New York (Academic Press)
\bibitem{} Tscharnuter W. M., 1971, Ap\&SS, 14, 11
\bibitem{} Wielen R., 1972, in Lecar M., ed., {\it Gravitational $N$-body
problem}, Proceedings of IAU Colloquium 10, Reidel
Publishing Company, Dordrecht
\bibitem{} Wielen R., 1974, in Mavridis L. N., ed., {\it Stars and the Milky Way System. Proceedings
of the First European Astronomocal Meeting, Athens, Sept. 4-9, 1972.}, Vol. 2, p. 326

\end{thebibliography}
\end{document}